\documentclass[12pt,a4paper]{article}
\usepackage{amsmath,amsfonts,amssymb,graphicx}

\makeatletter \@addtoreset{equation}{section} \makeatother

\begin{document}

\title{ Some  constraints on the Yukawa
parameters\\ in  the neutrino modification of the Standard Model
($\nu MSM$) and CP-violation}
\author{Volodymyr M. Gorkavenko\thanks{E-mail:
gorka@univ.kiev.ua}\,\, and Stanislav I. Vilchynskiy\thanks{E-mail: sivil@univ.kiev.ua}\\
\it \small Department of Physics, Taras Shevchenko National
University of Kyiv,\\\small\it 64 Volodymyrs'ka St., Kyiv, 01601,
Ukraine}
\date{}
\maketitle

\begin{abstract}
The equations connecting elements of the Yukawa matrix to elements
of the active neutrino mass matrix in the $\nu MSM$ theory (an
extension of the Standard Model by a singlet of three right-handed
neutrinos) was analyzed, and explicit relations for the ratio of the
Yukawa matrix elements through  elements of the active neutrino mass
matrix were obtained. This relation can be used for
 getting  more accurate constraints on the model
parameters. Particularly, with the help of the obtained results we
investigated CP-violating phase in the $\nu MSM$ theory. We
demonstrate that even in the case when elements of the active
neutrino mass matrix are real the baryon asymmetry can be generated
also.
\end{abstract}
\renewcommand{\theequation}{\arabic{section}.\arabic{equation}}

\section*{Introduction}
\setcounter{equation}{0}

The Standard Model (SM) \cite{SM} of the electroweak and the strong
interactions is one of the greatest successes of the physics. It
 describes correctly the processes with  participation of
elementary particles to energy scale $\sim 100$ GeV and for
individual processes to several TeV. It predicted a number of
particles, the overwhelming majority of which has been observed.
However, it is well known  that the SM does not account for several
phenomena in particle physics, astrophysics and cosmology. Namely:
the SM does not provide the dark matter  candidate; the SM does not
explain neutrino oscillations and the baryon asymmetry of the
Universe; the SM can not solve the strong CP problem in particle
physics, the primordial perturbations problem and the horizon
problem in cosmology, etc.

The solutions of the  above mentioned problems of the SM need some
new physics between the electroweak and the Planck scales. An
important challenge for the theoretical physics is to see if it is
possible to solve them using only the extensions of the SM below the
electroweak scale \cite{last}.

The SM is the renormalized theory which is  based on $SU(3)\times
SU(2) \times U(1)$ gauge group and  contains three generations
fermions. The left-handed components of fermions form the weak
isospin doublets relative to the  $ SU (2) $ group, and the
right-handed components of all fermions except neutrinos are
singlets of the weak isospin. The absence of the right-handed
neutrino fields in the SM is due to the fact that neutrinos  are
considered as massless particles. However, the recent experimental
discovery of the neutrino oscillation phenomenon \cite{PG,Strumia}
(transitions between neutrinos of different flavors) is a proof that
neutrinos have a nonzero mass. The current data show that the mass
squared differences of active neutrino $\Delta
m_{atm}^2=(2.5\pm0.2)\cdot 10^{-3}eV^2$ and $\Delta
m_{sol}^2=(8.0\pm0.3)\cdot 10^{-5}eV^2$.

One of the simplest and the most promising ways of modification of
the SM is an extension of the fermionic sector by adding the singlet
of right-handed neutrinos, which do not take part in the SM gauge
interactions\footnote{This is why these neutrinos are called sterile
neutrinos. The left-handed neutrinos of the SM are called active
neutrinos.}. The consequence of the existence of two distinct scales
$\Delta m_{atm}^2$ and $\Delta m_{sol}^2$ is that number of
right-handed neutrinos must be greater than or equal to two. The
introduction of only two right-handed neutrinos leads to the
emergence of 11 new parameters in the modified theory, which may be
able to explain the available experimental data on the oscillations
of active neutrinos. In this case the model predicts the existence
of two massive active neutrinos and one massless neutrino, which
does not contradict the available experimental data. However, the
extension of the SM by only two right-handed neutrinos does not
solve the remaining problems of the SM, in particular, does not
explain the nature of the dark matter \cite{Shap2,new}.

It turns out that introduction of  three right-handed neutrinos with
masses smaller than electroweak scale can provide explanations of
the experimental data in  particle physics, astrophysics, and
cosmology. This model (called $\nu MSM$) \cite{Shap2,Shap1} is the
simplest extension\footnote{Various possibilities to incorporate
neutrino masses in the theory are discussed, e.g., in
\cite{other1}.} of the SM, which can explain simultaneously some of
phenomena of the new physics without a new energy scale.
 Moreover, the parameters of the model (it contains 18
new parameters --  three Majorana neutrino masses, three Dirac
neutrino masses, six mixing angles and six CP-violating phases)
 can be chosen in such a way that the physics above
the electroweak  scale is not altered, while the following three
phenomena beyond the SM are explained. a) All  data on the neutrino
oscillations can be fitted. The smallness
  of the neutrino masses is explained by the Type I of the "see-saw"\, mechanism
  \cite{2}.
b) Parameters of  two heavier neutrinos can be chosen to
  allow baryogenesis. The masses of these particles
  can still be chosen below the electroweak scale.
c) The lightest sterile neutrino can be intensively produced in the early
  universe and have cosmologically long life-time. So, it might be a viable
  dark matter  candidate.

The  topical problem is the determination of the parameters of the
$\nu MSM$.  One of the purpose of this work is to obtain
restrictions on the values of the  Yukawa's matrix elements from the
$\nu MSM$ equations  that connect elements of the Yukawa's matrix
with elements of the active neutrinos mass matrix.

Since neutrinos in the SM are massless particles, the only source of
CP-symmetry violation in weak interactions is a single complex
element of the
 Kabbibo-Kobayashi-Maskawa matrix, which describes  the mixing between
 the
quarks of different generations.  Due to the existence of neutrinos
with nonzero masses
 it becomes possible to mix
 different generations of neutrinos. Moreover, in the $\nu
MSM$ the mix between active and sterile neutrino is possible too.
So, there is another possible source of generation of CP-symmetry
violation and the $\nu MSM$ theory has a much richer structure than
 the SM.

Another purpose of this work is to study  CP violation in the frame
of the $\nu MSM$ theory. As  was shown in \cite{Shap2}, effect of
the CP violation on the baryogenesis in the $\nu MSM$ theory can be
expressed through the CP-violating factor $\delta_{CP}$. This factor
includes the above mentioned  CP-violating phases. The baryon
asymmetry is proportional to $\delta_{CP}$. An applicable expression
for $\delta_{CP}$ was obtained in \cite{leptonic}. Investigation of
this expression is the second goal of the work.

The paper is organized as follows. The first section contains basic
relations and some results obtained in the $\nu MSM$. In the second
section of the paper  the  $\nu MSM$ equations are analyzed in
detail, and the explicit relations for ratio of the Yukawa matrix
elements through the elements of active neutrino mass matrix are
obtained. Using results of the second section the
 CP-violating factor is analyzed in the third section.

\section{The basic relations and some results\\ of the $\nu MSM$}

In the $\nu MSM$  \cite{Shap2,Shap1} the following terms  are added
to the Lagrangian of the SM (without taking into account the kinetic
terms):
\begin{multline}\label{sh1}
\mathcal L^{ad}=-F_{\alpha I}\bar L_\alpha \tilde\Phi
\nu_{IR}-\frac{M_{IJ}}{2}\bar
\nu_{IR}^c\nu_{JR}+h.c.=\\=-\frac{h+v}{\sqrt{2}}\bar\nu_{\alpha
L}F_{\alpha I}\nu_{IR}-\bar \nu_{IR}^c\frac{M_{IJ}}{2}\nu_{JR}+h.c.,
\end{multline}
where index  $\alpha=e,\mu,\tau$  corresponds to the active neutrino
flavors, indices  $I,J$ run from $1$ to $3$, $L_{\alpha}$ is for the
lepton doublet, $\nu_{IR}$ the field functions of the sterile
right-handed neutrinos, the superscript $\raisebox{-0.8em}{$"$}\!c"$
means charge conjugation, $F_{\alpha I}$ is for the new (neutrino)
matrix of the Yukawa constants, $M_{IJ}$  for the Majorana mass
matrix of the right-handed neutrinos, $\Phi$ for the field of the
Higgs doublet in the unitary gauge, ${\tilde\Phi}=i\sigma_2\Phi^*$,
$h$ is for the neutral Higgs field and the parameter $v$ determines
  minimum of the
Higgs  field potential
 $(v\cong247\,\,\mbox{GeV})$.

In the SM the mass of fermions are generated due to interactions of
fermions fields with scalar Higgs  field. The structure of the SM is
such that after spontaneous symmetry breaking the neutrino remains
to be a massless particle\footnote{There is a possibility to
introduce the mass of neutrino after the electroweak symmetry
breaking with help of the effective dimension 5 non-renormalizable
operator $\frac{\lambda_{ij}}\Lambda (L_i
\Phi)^T(L_j\Phi),\,\,\,i,j=e,\mu,\tau$, where $L$ is the SU(2)
lepton doublets, $\Lambda$ is the cutoff high-energy scale
\cite{Weinberg}. This operator breaks lepton number and can be
obtained from \eqref{sh1} by integrating out heavy right-handed
neutrinos.}. The SM does not contain Dirac mass term $\sim \bar
\nu_L\cdot \nu_R$ due to the absence in theory the right-handed
neutrinos,
 and the  Majorana mass term
$\sim \bar \nu_L^c\cdot\nu_L$ is forbidden by $SU(2)_L$ invariance.
The assumption about the existence of the right-handed neutrinos
leads to the appearance  of the Dirac as well as Majorana mass terms
in the Lagrangian, which in the general case have the form (see,
e.g., \cite{Belenki})
\begin{equation}\label{dm1}
\mathcal L^{DM}=-\left(\overline{(N_L)^c}
\,\frac{M^{DM}}2\,N_L+h.c.\right),
\end{equation}
where
\begin{equation}\label{dm2}
N_L=\left(%
\begin{array}{c}
  \nu_L \\
  \nu_R^c \\
\end{array}
\right)\!;\,\, N_L^c=\left(
\begin{array}{c}
  \nu_L^c \\
  \nu_R \\
\end{array}
\right)\!;\,\,
M^{DM}=\left(%
\begin{array}{cc}
  M_L&M_D{}^T \\
  M_D&M_R \\
\end{array} \right)\!.
\end{equation}
Comparing  mass terms in  \eqref{sh1} with \eqref{dm1} one can see
\begin{equation}\label{dm1a} M_L=0,\quad
M_D=F^+\frac{v}{\sqrt2}\,,\quad M_R=M^*,
\end{equation}
where $M,F$ are square matrix of the third order with elements
$F_{\alpha I}$ and $M_{IJ}$ (\ref{sh1}).

As it was shown in \cite{Shap2,Shap1},   the parameters of the $\nu
MSM$ can be chosen in such a way to simultaneously explain the
neutrino oscillations, the baryon asymmetry and to determine the
nature of the dark matter. This requires the existence of two
right-handed neutrinos with large practically the same masses
($\gtrsim$100 MeV) and one right-handed neutrino with a relatively
small mass\footnote{For the time being the allowed region for the
mass of the lightest sterile neutrino is $(1\div 50)$ KeV
\cite{last}.}.

In  zero approximation the extended Lagrangian  $\mathcal L_{\nu
MSM}$ is assumed to be
 invariant under $U(1)_e\times U(1)_\mu\times
U(1)_\tau$ transformations,  that provides preservation of the $e,
\mu, \tau$ lepton numbers separately as in the SM. In addition, it
is assumed that  two heavy sterile neutrinos interact with the
active neutrinos, but the third (lightest) sterile neutrino does not
interact\footnote{The lightest neutrino is the dark matter candidate
in the $\nu MSM$, just
 because it does not have to interact with other particles of the SM.}. This
 assumption can be realized by following matrix $M^{DM}$ \cite{Shap3}:
\begin{equation}\label{sh3}
M_L^{(0)}=0;\quad M_R^{(0)}=\left(%
\begin{array}{ccc}
  0 & 0 & 0 \\
  0 & 0 & M \\
  0 & M & 0 \\
\end{array}%
\right)\!,\quad M_D^{(0)+}=\frac{v}{\sqrt{2}}\left(%
\begin{array}{ccc}
  0 & h_{12} & 0 \\
  0 & h_{22} & 0 \\
  0 & h_{32} & 0 \\
\end{array}%
\right)\!,
\end{equation}
where $M$ is real and  $h_{12},$ $h_{22},$ $h_{32}$ -- are complex
quantities.

In this approximation  we have two massive right-handed neutrinos
with equal mass $M$, the third right-handed neutrino is massless,
and all active neutrinos have a zero mass, which is in contradiction
with observable data. To adjust it next small  terms are added to
the matrix $M_R$ and $M_D$ \cite{Shap3}:
\begin{equation}\label{sh4}
M_L^{(1)}=0;\,\, M_R^{(1)}=\left(\!\!
\begin{array}{ccc}
  m_{11}e^{-i \alpha} & m_{12} & m_{13} \\
  m_{12} & m_{22}e^{-i \beta} & 0 \\
  m_{13} & 0 & m_{33}e^{-i \gamma} \\
\end{array}%
\!\!\right);\,\, M_D^{(1)+}=\frac{v}{\sqrt{2}}\left(\!\!
\begin{array}{ccc}
  h_{11} & 0 & h_{13} \\
  h_{21} & 0 & h_{23} \\
  h_{31} & 0 & h_{33} \\
\end{array}%
\!\!\right)\!,
\end{equation}
where $m_{ij}$ ($m_{ij} \ll M$) are real, but elements of first and
second columns ($|h_{i1}| \ll |h_{k2}|$, $|h_{i3}| \ll |h_{k2}|$)
are in general complex elements.

These corrections break $U(1)_e\times U(1)_\mu\times
U(1)_\tau$-symmetry, lead to the appearance of a small mass of the
third right-handed neutrino, and remove the mass degeneracy for  two
heavy right-handed neutrinos. It ensures the appearance of extra
small masses of the active  neutrinos  and nonzero mixing angles
among them.

As is well known  \cite{Belenki}, the mass part of the Lagrangian
\eqref{dm1} can be  diagonalized by the transition from the basis of
the gauge functions $N_L$ to the basis of the mass
 functions  $n_L$ using  an unitary matrix $V$, namely
$N_L=Vn_L$, so
\begin{equation}\label{dm8}
\bar N_L=\bar n_LV^+;\quad N_L{}^c=(V^+)^Tn_L^c; \quad \overline
N_L{}^c=\overline{(n_L)^c}V^T,
\end{equation}
where $V$ $(6\times6)$ is a product of  two matrices $V=WU$.

 The  $W$ matrix is introduced for the block
diagonalization of  the $M^{DM}$ matrix \cite{BelGi}. The explicit
form of the $W$ matrix  can be approximately found in the
"see-saw"\, approach due to the smallness of matrix
 $\varepsilon=M_R^{-1}M_D\ll1$:
\begin{equation}\label{ss3}
W=\left(
\begin{array}{cc}
  1-\frac{1}{2}\varepsilon^+\varepsilon & \varepsilon^+ \\
  -\varepsilon & 1-\frac{1}{2}\varepsilon\varepsilon^+ \\
\end{array}%
\right).
\end{equation}
In this approximation the result of the block diagonalization has
the form
\begin{equation}\label{ss4}
 W^TM^{DM}W=\left(%
\begin{array}{cc}
  -M_D^TM_R^{-1}M_D & 0 \\
  0 & M_R \\
\end{array}%
\right)=\left(%
\begin{array}{cc}
  M_{light} & 0 \\
  0 & M_{heavy} \\
\end{array}%
\right),
\end{equation}
where matrices $M_{light}$ and  $M_{heavy}$ are the mass matrices of
the active and the sterile neutrinos  correspondingly. Note that the
elements of the matrix $M_{light}$ and accordingly the masses of the
active neutrinos are completely determined by elements of the
matrices $M_D$ and $M_R$.

The $U$  matrix has the form
\begin{equation}\label{ss4a}
U=\left(%
\begin{array}{cc}
  U_1 & 0 \\
  0 & U_2 \\
\end{array}%
\right),
\end{equation}
where matrices $U_{(1,2)}$ (each of them are $(3\times 3)$ matrix)
are chosen for the diagonalization of block matrix $W^TM^{DM}W$
\begin{multline}\label{ss5}
m=diag(m_1,m_2,\ldots m_6)=V^TMV=\\=U^TW^TMWU=\left(
\begin{array}{cc}
  U_1^TM_{light}U_1 & 0 \\
  0 & U_2^TM_{heavy}U_2 \\
\end{array}%
\right)\!.
\end{multline}

There is a standard parametrization
 \cite{PG} for $U_{(1,2)}$:
\begin{multline}\label{ss8}
U_{(1,2)}=\left(\!\!
\begin{array}{ccc}
  c_{12}c_{13}  & c_{13}s_{12} &  s_{13}e^{-i \delta}  \\
    -s_{12}c_{23}-c_{12}s_{23}s_{13}e^{i \delta}  & c_{12}c_{23}-s_{12}s_{23}s_{13}e^{i \delta} & s_{23}c_{13}  \\
  s_{12}s_{23}-c_{12}c_{23}s_{13}e^{i \delta} & -c_{12}s_{23}-s_{12}c_{23}s_{13}e^{i \delta} & c_{23}c_{13}  \\
\end{array}%
\!\!\right)\times\\\times\left(\!\!
\begin{array}{ccc}
  e^{i \alpha_1 /2} & 0 & 0 \\
  0 & e^{i \alpha_2 /2} & 0 \\
  0 & 0 & 1 \\
\end{array}%
\!\!\right)\!\!,
\end{multline}
where $c_{ij}=\cos \theta_{ij}$, $s_{ij}=\sin \theta_{ij}$,
$\theta_{12},\theta_{13},\theta_{23}$ are the three mixing angles;
$\delta$ is the Dirac phase, and $\alpha_1, \alpha_2$ are the
Majorana phases. The angles  $\theta_{ij}$ can be  in the region
$0\leq\theta_{ij}\leq\pi/2$, phases $\delta,\alpha_1, \alpha_2$ vary
from  $0$ to $2\pi$. Each of the matrices $U_{(1)}$ and $U_{(2)}$
contains its own, independent angles and phases.

Thus, the determination of the masses of active and sterile
neutrinos is reduced to the dia\-go\-nali\-za\-tion of the matrix
\eqref{ss4}, where diagonalization can be carried out separately for
the matrices $M_{light}$ and $M_{heavy}$. Since the matrix
$M_{light}$ and $M_{heavy}$ are not Hermitian, it is more
appropriate to find eigenvalues of Hermitian matrices
$M_{light}^+M_{light}$ and $M_{heavy}^+M_{heavy}$ by solving
corresponding equations. The found eigenvalues are the square of
eigenvalues of the matrices
 $M_{light}$ and $M_{heavy}$.

In the approximation  when the elements of the first column of the
Yukawa matrix are neglected and  $M\gg m_{ij}$,
 the mass of the lightest active neutrino is zero. The nondiagonal mass matrix of the active neutrinos has the form
\cite{Shap3}
\begin{equation}\label{6}
[M_{light}]_{\alpha \beta }=\eta( h_{\beta 3}h_{\alpha 2}+ h_{\alpha
3}h_{\beta 2}),\quad \mbox{where}\quad\eta=v^2/2M
\end{equation}
and eigenvalues
\begin{equation}\label{sh9}
m_{2,3}=\eta(F_2 F_3 \pm |h^+ h|_{23}),
\end{equation}
where $F_2^2=[h^+h]_{22}$, $ F_3^2=[ h^+ h]_{33}$, $F_2
F_3=M(m_2+m_3)/v^2$.

On the other hand, the elements of the matrix $M_{light}$ are
defined by masses and mixing matrix $U_{(1)}$ of the active
neutrinos \cite{Belenki}:
\begin{equation}\label{13}
[M_{light}]_{\alpha \beta
}=m_1U_{(1)i1}^*U_{(1)j1}^*+m_2U_{(1)i2}^*U_{(1)j2}^*+m_3U_{(1)i3}^*U_{(1)j3}^*.
\end{equation}
The elements of the mixing matrix $U_{(1)}$ are known (unfortunately
with a considerable inaccuracy) from the neutrino oscillation
experiments (see, e.g., \cite{PG,Strumia}). Parameters of the matrix
$U_{(1)}$ \eqref{ss8} are presented in  Tab.1.

\begin{center}
\begin{tabular}{c c}\hline
 central value & 99$\%$ confidence interval \\
\hline  $|\Delta m^2_{12}| = (8.0\pm0.3)\cdot10^{-5}\,eV^2$&
 $(7.2 - 8.9)\cdot10^{-5} eV^2$\\
 $|\Delta m^2_{23}| = (2.5
\pm 0.2) \cdot 10^{-3} eV^2$& $(2.1 -
3.1) 10^{-3} eV^2$\\
 $tan^2 \theta_{12} = 0.45 \pm 0.05$& $30^0 < \theta_{12} < 38^0$\\
 $sin^2 2\theta_{23} = 1.02 \pm 0.04$& $ 36^0 < \theta_{23} < 54^0$\\
 $sin^2 2\theta_{13} = 0 \pm 0.05$&$ \theta_{13} < 10^0$\\
\hline
\end{tabular}
\end{center}

{\small Table 1. Experimental constraints on the parameters of
active neutrinos.}

\section{The ratio of the Yukawa matrix elements\\ in the $\nu MSM$}

The system of equations \eqref{6} connects elements of the second
and the third columns of  the Yukawa sterile neutrinos matrix with
elements
 of the active neutrinos matrix and has  infinite number of
solutions. Indeed, the replacement of  $h_{i2}$ to $zh_{i2}$ and $
h_{i3}$ to $ h_{i3}/z$ ($z$ is an arbitrary complex number) does not
change the system \eqref{6}.

But the system of equations \eqref{6} has a good solutions for the
ratio of the Yukawa matrix elements. Indeed, if we take ${ h}_{13},{
h}_{23}$, and ${ h}_{33}$ from the equations for the diagonal
elements of $M_{light}$ and substitute it into equations for the
off-diagonal elements of $M_{light}$ we get
\begin{equation}\label{10}
\left\{\begin{array}{c}
\vspace{0.5em}M_{12}=\displaystyle{\frac12\left(M_{22}\frac{h_{12}}{h_{22}}+M_{11}\frac{h_{22}}{h_{12}}\right)}\\
\vspace{0.5em}M_{13}=\displaystyle{\frac12\left(M_{33}\frac{h_{12}}{h_{32}}+M_{11}\frac{h_{32}}{h_{12}}\right)}\\
M_{23}=\displaystyle{\frac12\left(M_{33}\frac{h_{22}}{h_{32}}+M_{22}\frac{h_{32}}{h_{22}}\right)}
\end{array}\right.
\end{equation}
with the obvious solutions
\begin{equation}\label{a9}
\left\{\begin{array}{c}
\vspace{0.5em}A_{12}=\displaystyle{\frac{M_{12}}{M_{22}}\left(1\pm\sqrt{   1-\frac{M_{11}M_{22}}{M_{12}{}^{2}} }  \right)}\\
\vspace{0.5em}A_{13}=\displaystyle{\frac{M_{13}}{M_{33}}\left(1\pm\sqrt{1-\frac{M_{11}M_{33}}{M_{13}{}^{2}} }  \right)}\\
A_{23}=\displaystyle{\frac{M_{23}}{M_{33}}\left(1\pm\sqrt{1-\frac{M_{22}M_{33}}{M_{23}{}^{2}}}\right)}
\end{array}\right.
\end{equation}
for the ratio of the second column elements of the Yukawa matrix
\begin{equation}\label{a7}
A_{12}=h_{12}/h_{22};\qquad A_{13}=h_{12}/h_{32};\qquad
A_{23}=h_{22}/h_{32}.
\end{equation}

The ratio of the third column elements of the Yukawa matrix can be
easily obtained from the  equations for the diagonal elements of
$M_{light}$:
\begin{equation}\label{dop}
\frac{h_{23}}{h_{13}}=A_{12}\frac{M_{22}}{M_{11}};\quad
\frac{h_{33}}{h_{13}}=A_{13}\frac{M_{33}}{M_{11}}.
\end{equation}

 So, in the approximation when \eqref{6} is valid the ratio of
the Yukawa matrix elements depends only on the active neutrino mass
matrix.

Though formally there are eight different choices for the solutions,
 only four are independent. For example, if we fix the sign before
the square roots in the expressions for $A_{12}$ and $A_{13}$ then
$A_{23}$ is unambiguously determined by the relation
\begin{equation}\label{a10}
A_{23}=A_{13}/A_{12}.
\end{equation}

It should be emphasized that solutions (\ref{a9}) do not allow one
to find elements of the Yukawa matrix but only its ratio\footnote{It
can be shown that our results \eqref{main3}, \eqref{main4} coincide
with results of \cite{leptonic} where the ratios of the elements
were obtained in the particular case $\theta_{13}\rightarrow 0$,
$\theta_{23}\rightarrow\pi/4$.}:
 \begin{equation}\label{main3}
\frac{\{h_{12}; h_{22}; h_{32}\}}{F_2}=\frac{e^{i\cdot
arg(h_{12})}}{\sqrt{1+\left|{A_{12}}\right|^{-2}+\left|{A_{13}}\right|^{-2}}}\left\{1;{A_{12}}^{-1};{A_{13}}^{-1}\right\},
\end{equation}
 \begin{equation}\label{main4}
\frac{\{  h_{13};   h_{23};   h_{33}\}}{ F_3}=\frac{e^{i\cdot arg(
h_{13})}}{\sqrt{1+\left|A_{12}\frac{M_{22}}{M_{11}}\right|^2+\left|A_{13}\frac{M_{33}}{M_{11}}\right|^2}}\left\{1;
A_{12}\frac{M_{22}}{M_{11}}; A_{13}\frac{M_{33}}{M_{11}}\right\},
\end{equation}
where phases of $h_{12}$, $  h_{13}$ are connected by condition
\begin{equation}\label{main5}
arg(h_{12})+arg(  h_{13})=arg(M_{11})
\end{equation}
and $A_{12},A_{13},M_{11},M_{22},M_{33}$ are definitely expressed
via parameters of the active neutrino mass matrix. Since the system
of equations \eqref{6} is written in the approximation
\mbox{$m_1=0$}, the phase $\alpha_1$ is excluded from all
expressions (see \eqref{ss8} and \eqref{13}). In this approximation
only seven parameters of active neutrinos are used: two mass $m_2,
m_3$; three mixing angles $\theta_{12},\theta_{13},\theta_{23}$; one
Dirac $\delta$ and one Majorana   $\alpha_2$ CP-violating phases.

To know straight  values of the Yukawa matrix elements we have to
know in addition two arbitrary parameters\footnote{ Only the
relation between absolute values of the elements of the second and
third columns of the Yukawa matrix $\epsilon=F_3/F_2$ was used as an
additional relation in \cite{Shap3} . It allows one to find out only
the absolute values of the Yukawa's elements.} of Yukawa matrix,
e.g., absolute value and phase of an arbitrary Yukawa matrix element
or quantities $\xi$ and $\epsilon$ from \eqref{epsilon}.

It can be shown that (for fixed values of the active neutrino
parameters) there are only two choices\footnote{This assertion is
always true, except the special case of the parameters when at least
one radical expression in \eqref{a9} is zero.} for placing of the
signs in the expressions
 for $A_{12},A_{13},A_{23}$  \eqref{a9} which are not inconsistent
with condition \eqref{a10}. These two variants are distinguished
from each other by simultaneous replacement of the sign in front of
square roots in the expressions for $A_{12},A_{13}, A_{23}$. It  can
be shown that such replacement of the signs leads to interchanging
and conjugating of the relation between elements of the second and
the third columns of the Yukawa matrix, notably
$h_{22}/h_{12}\leftrightarrow h^*_{23}/h^*_{13}$,
$h_{32}/h_{12}\leftrightarrow h^*_{33}/h^*_{13}$.

The  solutions of (\ref{6}) can be analyzed numerically. Really,
using condition \eqref{a10} for setting correct signs in \eqref{a9},
we can straightly find the ratio of elements of the second
$\left(|h_{22}/h_{12}|=A_{12}^{-1};
|h_{32}/h_{12}|=A_{13}^{-1}\right)$ and the third
$\left(|h_{23}/h_{13}|\!=\!A_{12}\frac{M_{22}}{M_{11}};
|h_{33}/h_{13}|\!=\!A_{13}\frac{M_{33}}{M_{11}}\right)$ columns of
the Yukawa matrix for every fixed point in the space of values
$m_2,m_3,$ $\theta_{12}$, $\theta_{13}$, $\theta_{23},$
$\alpha_2,\delta$.

In the case of the normal hierarchy  the masses of active neutrinos
increase with increasing their numbers ($m_1<m_2<m_3$). In the $\nu
MSM$ $m_1=0$, so we take central values $m_2=\sqrt{|\Delta
m_{12}^2|}=0.009\,eV$, $m_3=\sqrt{|\Delta m^2_{23}|}=0.05\,eV$. The
phases $\delta$, $\alpha_2$  vary in range from  $-\pi$ to $\pi$,
the angles $\theta_{12},\theta_{13},\theta_{23}$ vary in accordance
with Tab.1. With help of the numerical analysis we found the minimum
and the maximum values for the ratio of elements of the second
column of the Yukawa matrix and obtained
$0.65\lesssim\left|h_{32}/h_{12}\right|\lesssim 24.2$,
$1.4\lesssim\left|h_{22}/h_{12}\right|\lesssim 29.6$. Also, we found
values of ($|h_{22}/h_{12}|$; $|h_{32}/h_{12}|$) and
($Arg[h_{22}/h_{12}]$; $Arg[h_{32}/h_{12}]$) for $10^4$ points in
the space of values $\theta_{12}$, $\theta_{13}$, $\theta_{23},$
$\alpha_2,\delta$. Results of calculations  are demonstrated in
Fig.1.

\phantom{hhh}

\begin{center}
\mbox{\includegraphics{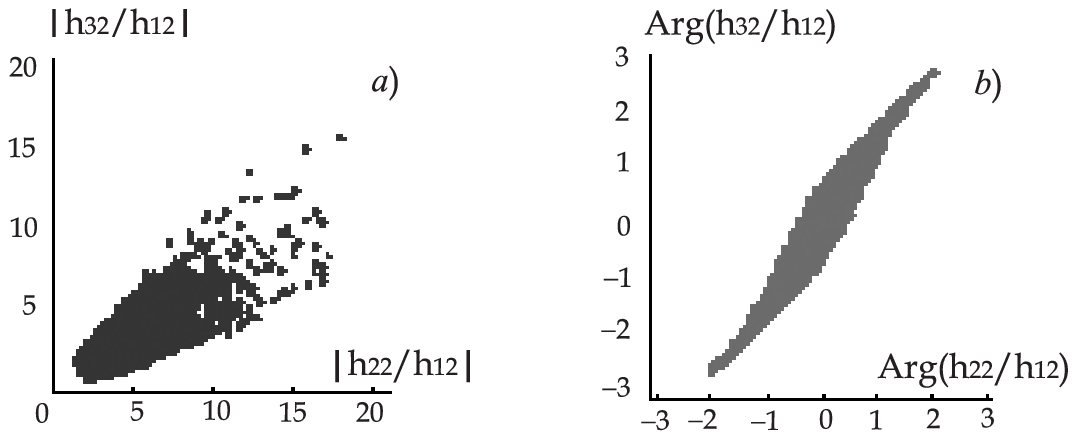}}\label{figure}
\end{center}
{\noindent \small Fig.1 The ratio of modules \textit{a}) and
difference of phases \textit{b}) of the second column elements of
the Yukawa matrix in the case of the normal hierarchy.}

\phantom{hhh}

It should be noted that if one arbitrary takes a point in the space
of values $\theta_{12}$, $\theta_{13},\theta_{23},\alpha_2,\delta$
there is a strong probability that $\left|h_{32}/h_{12}\right|$ and
$\left|h_{22}/h_{12}\right|$ will lie in range from 1 to 10 and  it
is improbable that ratio of the Yukawa couplings will be greater
than ten. The values of the phase differences
$Arg\left[h_{32}/h_{12}\right]$ and $Arg\left[h_{22}/h_{12}\right]$
lie not in full region $[-\pi,\pi]$ but only in a closed compact
region (Fig.1,\textit{b}).

In the case of the inverse hierarchy the masses of active neutrinos
increase with reducing their numbers ($m_1>m_2>m_3$) but the $\nu
MSM$ predicts $m_1=0$. In order to conform it we can swap the mass
state $m_3$ and $m_1$ with help of additional rotation via unitary
anti-diagonal matrix $\tilde U$: $U_{(1)}\rightarrow U_{(1)}\tilde
U$ where $U_{(1)}$ is the mixing matrix of the active neutrinos
\eqref{ss8}. Assuming $m_1=0$, we get central values
$m_2=m_3=\sqrt{|\Delta m^2_{23}|}=0.05\,eV$ and found the same
quantities for the same range of values $\theta_{12}$,
$\theta_{13}$, $\theta_{23},\alpha_2,\delta$ like in the case of
normal hierarchy. In this case the ratio of elements of the second
column of the Yukawa matrix lies in range
$0\leq\left|h_{32}/h_{12}\right|\lesssim 3.2$,
$1.1\lesssim\left|h_{22}/h_{12}\right|\lesssim 4.3$. The points in
the space of values $\theta_{12}$, $\theta_{13}$, $\theta_{23},$
$\alpha_2,\delta$  are demonstrated by Fig.2.

In this case the boundary  great values of the elements ratio are
improbable too. The ratio $\left|h_{32}/h_{12}\right|$ can be equal
to zero because of $\left|h_{32}/h_{12}\right|=A_{12}^{-1}\sim
M_{12}$ and $M_{12}$ can be equal zero in allowed range of
parameters (Tab.1) under condition $m_1=m_2$. So, in contrast to the
case of the normal hierarchy, elements $|h_{i2}|$ can be of
different order of magnitude. Similarly to the case of the normal
hierarchy, the phase differences $Arg\left[h_{32}/h_{12}\right]$ and
$Arg\left[h_{22}/h_{12}\right]$ lies in closed compact region
(Fig.2,\textit{b}).

\phantom{hhh}

\begin{center}
\mbox{$\!\!\!\!\!\!\!\!\!\!$\includegraphics{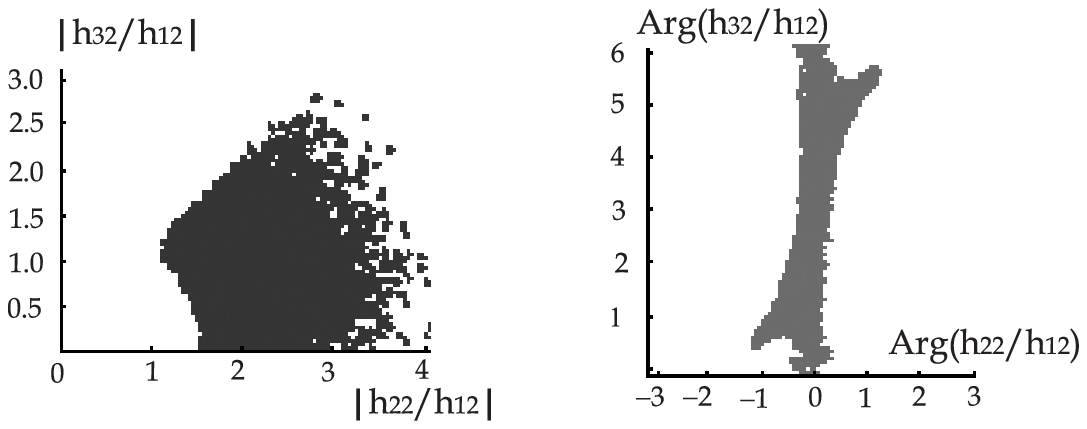}}\label{figure}
\end{center}

{\noindent \small Fig.2 The ratio of modules \textit{a}) and
difference of phases \textit{b}) of the second column elements of
the Yukawa matrix in the case of the inverse hierarchy.}

\phantom{hhh}

In the case of normal and inverse hierarchies the graphical
representation of ratio  of the third column elements
($\left|h_{33}/h_{13}\right|$; $\left|h_{23}/h_{13}\right|$)  is
identical to Fig.1,2(\textit{a}). The corresponding representation
of the phase differences ($Arg\left[h_{33}/h_{13}\right];
Arg\left[h_{23}/h_{13}\right]$) is identical to Fig.1,2(\textit{b})
but in the region $[0;2\pi]$.

The obtained constraints on the possible values of ratio of the
Yukawa sterile neutrinos matrix elements are determined by the
current data on active neutrino parameters. When the inaccuracy of
the active neutrino parameters will decrease the constraints will be
improved too. Obtained results can be useful for getting more
accurate predictions of the $\nu MSM$ theory particularly for the
investigation of the baryon asymmetry.

As one can see, there are four independent variants of solutions of
\eqref{6}.  It should be noted that even in the case when the masses
and mixture angles of active neutrinos are known exactly we will be
not able to say which of these  variants of the solutions are
realized. To known it we  need the values
 of phases of the active neutrinos matrix $U_1$. For  fixed values of the all active
 neutrino matrix
parameters the quantity of the possible solutions reduces to two
that corresponds to interchanging and conjugating of the relation
between elements of the second and the third columns of the Yukawa
matrix. It is closely related to the symmetry of  \eqref{6}
 under replacing the elements of the second column of the Yukawa
matrix by elements of the third column.  In this case two  variants
of solutions in principle can be distinguished experimentally with
help of measurement of the mixture angles between active and sterile
neutrinos.

\section{CP-violation in the  $\nu
MSM$}

As is known \cite{Saxar},  in order that the baryon asymmetry
generated from the initial charge symmetric state of the hot
Universe, the next conditions must be satisfied simultaneously: a)
baryon number non-conservation; b) C- and CP-violation; c)
deviations from  thermal equilibrium.

In the field theory the Lagrangian under CP-transformations turns
into the Lagrangian with  complex conjugated couplings. If theory
contains couplings with unremovable phases, then this theory is not
CP-invariant.  The unique  source of CP-violation in the SM is one
complex  element in the Cabibbo-Kobayashi-Maskawa mixing matrix of
 quarks. In the $\nu MSM$ theory, owing to massiveness of
neutrinos, there is mixing among different generations of neutrinos
 and therefore there is additional possible source of CP-symmetry
 violation. This theory has a
 possibility of the baryon asymmetry generation  due to existence of
 CP-violating oscillations of active neutrinos into sterile. Such
 oscillations change the full lepton number of a system and create
 lepton asymmetry that transforms into the baryon asymmetry with the help
 of the electroweak sphalerons \cite{last,CP}.

To analyze CP-symmetry violation in the $\nu MSM$ we use relation
for the CP-violating factor \cite{leptonic}
\begin{equation}\label{cp1}
\delta_{CP}=\frac{1}{F_2^{6}}\left [Im(h^+h)_{23}\sum_\alpha
(|h_{\alpha 2}|^4-|h_{\alpha
3}|^4)-(F_2^2-F_3^2)\sum_\alpha(|h_{\alpha 2}|^2+|h_{\alpha
3}|^2)Im[h_{\alpha 2}^*h_{\alpha 3}]\right]\!.
\end{equation}
The solutions  (\ref{a9}) allow us to express the last relation
through the parameters of the active neutrinos. We get
\begin{multline}\label{cp12}
\delta_{CP}(\xi,\varepsilon)=|M_{11}|^{-1}C^{-3}
\left[\varepsilon\left(Im\left [e^{-2i\xi}A\right]
B-CD\right)+\varepsilon^3(C_1D-CD_1)+\right.\\\left.+\varepsilon^5\left
(C_1D_1-B_1Im\left [e^{-2i\xi}A\right]\right )\right],
\end{multline}
where there is dependence on only two parameters of the Yukawa
matrix
 \begin{equation}\label{epsilon}
 \xi=arg[h_{12}],\qquad\varepsilon=|h_{13}/h_{12}|=\epsilon\sqrt{C/C_1},
\end{equation}
and following notations are used:
$$
\begin{array}{l}
{\displaystyle \epsilon=F_3/F_2;\,\,A=
M_{11}+\frac{A_{12}}{A_{12}^*}M_{22}+\frac{A_{13}}{A_{13}^*}M_{33}};\\
{\displaystyle
\vphantom{\int}B=1+|A_{12}|^{-4}+|A_{13}|^{-4};\qquad\qquad\,\,\,
C=1+|A_{12}|^{-2}+|A_{13}|^{-2}};\\
{\displaystyle B_1=1+\left |A_{12}\frac{M_{22}}{M_{11}} \right
|^4+\left |A_{13}\frac{M_{33}}{M_{11}} \right |^4;\quad C_1= 1+\left
|A_{12}\frac{M_{22}}{M_{11}} \right |^2+\left|
A_{13}\frac{M_{33}}{M_{11}}\right|^2 };\vspace{0.3em}\\
{\displaystyle D=Im\left[e^{-2i\xi}M_{11}\right]+|A_{12}|^{-2}Im
\left[ e^{-2i\xi}\frac{A_{12}}{A_{12}^*}M_{22}\right]
+|A_{13}|^{-2}Im \left[ e^{-2i\xi}\frac{A_{13}}{A_{13}^*}M_{33}\right]};\vspace{0.3em}\\
{\displaystyle D_1=Im\left [e^{-2i\xi}M_{11}\right]+\left
|\frac{M_{22}}{M_{11}}A_{12}\right|^2\!Im \left
[e^{-2i\xi}\frac{A_{12}}{A_{12}^*}M_{22}\right]+\left
|\frac{M_{33}}{M_{11}}A_{13}\right|^2\!Im \left
[e^{-2i\xi}\frac{A_{13}} {A_{13}^*}M_{33}\right]}.
\end{array}
$$

Numerical analysis of the relation \eqref{cp12} confirms the general
properties of the  CP-violating factor \eqref{cp1} obtained in
\cite{leptonic}:

\noindent 1) sign of the CP-violating factor and correspondingly the
sign of the baryon asymmetry can not be determined by only the
elements of the active neutrino matrix;

\noindent 2) if $\epsilon\rightarrow 0$, then
$\delta_{CP}\sim\epsilon$ and also tends to zero;

\noindent3) the CP-violating factor can not be equal zero
\footnote[9]{For some particular values of  mixing angles and phases
$\delta_{CP}$ may be zero. This values of the parameters can be
found numerically.} when $\epsilon= 1$;

\noindent 4) the CP-violating factor can not be equal zero${^9}$
when $\theta_{13} = 0$ and $\theta_{23} = \pi/4$;

\noindent 5) in the case of inverse hierarchy the CP-violating
factor can not be equal zero${^9}$ when $m_1 = m_2$, $\theta_{13} =
0$, $\theta_{23} = \pi/4$.

The range of values of all parameters in the relation  (\ref{cp12})
is known\footnote[10]{The bounds of values of the mixing angles are
defined by Tab.1, the phases are in the region $[0,2\pi]$,
$7\cdot10^{-5}<\epsilon<1$ \cite{leptonic}.}. This allows one to
estimate bounds of possible values of the CP-violating factor. For
the case of normal hierarchy we have $|\delta_{CP}|\lesssim0.27$,
for the case of inverse --- $|\delta_{CP}|\lesssim0.08$.

As one can see from  \eqref{cp1} the CP-violating factor may be
nonzero only in the case when Yukawa matrix elements have an
imaginary part. The solutions (\ref{a9}) allow one to consider the
possibility of the baryon asymmetry generation in the case when
mixing matrix $U_{(1)}$ and, correspondingly to (\ref{13}), the mass
matrix of the active neutrino $M_{light}$ are real. It is easy to
see that for
 real matrix $U_{(1)}$ \eqref{ss8} the following minors of the matrix
$M_{light}=U_{(1)}^*m U_{(1)}^+$ (here $m=diag(0,m_2,m_3)$) are not
negative:
\begin{equation}\label{4}
\left\{\begin{array}{l}
M_{11}M_{22}-M_{12}^2=m_2m_3(\sin\theta_{13}\cos\theta_{12}\cos\theta_{23}-\sin\theta_{12}\sin\theta_{23})^2\geq0; \\
M_{11}M_{33}-M_{13}^2=m_2m_3(\sin\theta_{13}\cos\theta_{12}\sin\theta_{23}+\sin\theta_{12}\cos\theta_{23})^2\geq0;\\
M_{22}M_{33}-M_{23}^2=m_2m_3\cos^2\theta_{13}\cos^2\theta_{12}\geq0.
\end{array}\right.
\end{equation}
If the values of the mixing angles are defined by Tab.1 all the
above minors are positive. In this case the ratio of Yukawa's
elements (\ref{a7}), (\ref{dop}) and, consequently, the Yukawa's
elements are complex numbers. So, we can expect that CP-violating
factor can be nonzero.

The direct numerical analysis of the CP-violating factor
\eqref{cp12} confirms this assumption. Furthermore, if one takes
arbitrary a point in the space of values $\theta_{12}$,
$\theta_{13},\theta_{23},\xi,\varepsilon$ ($\delta=\alpha_2=0$)
there is a strong probability that the CP-violating factor will be
nonzero\footnote[11]{The equation  for zero CP-violating factor
\eqref{cp12} has a fine-tuning solutions in space of the mentioned
parameters.}. Thus, even in the case when active neutrino mass
matrix $M_{light}$ is real the electroweak generation of the baryon
asymmetry can be realized also.

The CP-violating factor is a complicated function of parameters of
the active neutrino matrix and the Yukawa matrix. The analysis of
this factor is accentuated by  fact that for fixed point in space of
active neutrino parameters one needs to take appropriate variant of
solutions of equations \eqref{6}. The investigation of manifestation
of the different variants of solutions \eqref{a9} on the
CP-violating factor is an interesting  task for future.

\section{Conclusions}

The $\nu MSM$ is the minimal neutrino modification of the SM that
can explain simultaneously neutrino oscillations, generation of the
baryon asymmetry, and the nature of dark matter. There are strong
conditions on the parameters of the $\nu MSM$ that
 can be experimentally checked. For the time being observable
data, obtained from the missions XMM-Newton, Chandra, INTEGRAL,
Suzaku, reveal no signs of existence of the sterile neutrino in
predicted by the $\nu MSM$ and instrumentally allowed region  (see
\cite{last} and references there). But new investigations are planed
(for example, project Xenia \cite{Xenia}) that will continue to
inspect theoretically allowed region of the model parameters.

Obtained in this paper exact solutions of the $\nu MSM$ equations
connect elements of the Yukawa matrix with  elements of the active
neutrino mass matrix and will be useful for analysis, data
processing, and getting more accurate constraints on the model
parameters.

The analysis of the ratio  of the Yukawa matrix elements
demonstrates that in the case of normal hierarchy elements of second
($h_{i2}$) and third ($h_{i3}$) columns are  the same order of
magnitude. But in the case of inverse hierarchy the magnitudes of
elements can considerably vary from each other  in the column.

CP-violating phase in the SM is  parameter of the
Cabibbo-Kobayashi-Maskawa matrix and it is known from observable
data. In contrast to it the CP-violating factor $(\delta_{CP})$ in
the $\nu MSM$ is effective parameter that is present in the
expression for the baryon asymmetry. Therefore the CP-violating
factor in the $\nu MSM$ has a sophisticated structure as a function
of the Yukawa matrix parameters.

 Obtained solutions (\ref{a9}) allow one to get expression for the
CP-violating factor (\ref{cp1}) through the parameters of the active
neutrinos and two parameters of the Yukawa matrix. Due to this the
bounds of possible values of the CP-violating factor were estimated.
It should be noted that in the case of the inverse hierarchy the
maximum of $|\delta_{CP}|$ is considerably smaller (fourfold) as
compared to the case of normal hierarchy.

As is known, the phases of the active neutrino mixing matrix can not
be measured for the time being. It was shown that in any case (even
if these phases are zero) the CP-violating factor can be nonzero and
the baryon asymmetry  generation is possible. The fact of the matter
is that  the elements of the Yukawa matrix are complex when the
elements of the active neutrino mixing matrix are real.

As it was mentioned above the $\nu MSM$ provides a candidate for
dark matter particle with mass $M_1$ in the range  (1$\div$50) KeV.
It is the lightest sterile neutrino that is produced due to the
resonant active-sterile neutrino oscillations in the presence of
lepton asymmetry \cite{leptonic,Shi}. It requires the high mass
degeneracy of two other heavier neutrinos and leads to the
fine-turning problem \cite{leptonic,Roy}.

It should be noted that modifications of the $\nu MSM$ can provide
other production mechanisms of the lightest sterile
 neutrino due to interactions with other new particles \cite{other} or primordial
  Higgs-inflation \cite{Higgs}.
It applies some additional constraints on the parameters of the $\nu
MSM$ (see, also, \cite{Roy})
 and changes the mass range of the lightest sterile neutrino.
The change of only $M_1$  has no action on the results of the
present work
 but the additional terms in the Lagrangians of such theories can modify the equations \eqref{6} that should be
taken into account  for  investigations of the constraints in such
theories.

We would like to thank A. Boyarsky, O. Ruchayskiy, D. Iakubovskyi
and M. Shaposhnikov for the idea of treating this subject, and for
useful comments and discussions. This work has been supported by the
Swiss Science Foundation (grant SCOPES 2010-2012, No.
IZ73Z0\_128040).

\phantom{cghjcgj}

The final publication is available at www.epj.org

\bibliographystyle{unsrt}

\end{document}